\documentclass[amsmath,twocolumn,prl,aps]{revtex4-1}
\usepackage{amsthm,amsfonts,graphicx,verbatim, color}
\usepackage{bm}
\usepackage[T1]{fontenc}
\usepackage[utf8]{inputenc}

\newcommand{\mb}[1]{\ensuremath{\mathbf #1}}

\newcommand{\eq}[1]{Eq.~\eqref{eq:#1}}

\newcommand{\Psb}{\ensuremath{\mathcal{P}_{\text{sb}}}}
\newcommand{\PLLL}{\ensuremath{\mathcal{P}_{\text{LLL}}}}

\graphicspath{{figs/}}

\begin{document}

\title{Many-body localization in Landau level subbands}

\author{Akshay Krishna${}^1$, Matteo Ippoliti${}^2$, and R. N. Bhatt${}^1$}
\affiliation{${}^1$Department of Electrical Engineering and ${}^2$Department of Physics, Princeton University, Princeton NJ 08544, USA}

\begin{abstract}
We explore the problem of localization in topological and non-topological nearly-flat subbands derived from the lowest Landau level, in the presence of quenched disorder and short-range interactions.
We consider two models: a suitably engineered periodic potential, and randomly distributed point-like impurities.
We perform numerical exact diagonalization on a torus geometry and use the mean level spacing ratio $\langle r \rangle$ as a diagnostic of ergodicity.
For topological subbands, we find there is no ergodicity breaking in both the one and two dimensional thermodynamic limits.
For non-topological subbands, in constrast, we find evidence of an ergodicity breaking transition at finite disorder strength in the one-dimensional thermodynamic limit.
Intriguingly, indications of similar behavior in the two-dimensional thermodynamic limit are found, as well.
This constitutes a novel, \emph{continuum} setting for the study of the many-body localization transition in one and two dimensions.
\end{abstract}

\maketitle

The problem of electron localization in the quantum Hall regime has a rich history.
The realization that the Hall conductance is robust to impurities\cite{Prange1981} paved the way for our understanding of the integer quantum Hall plateau transition as a problem of electron localization in a system with a diverging localization length protected by a topological invariant \cite{ Chalker1988, Huo1992, Huckestein1995}.
Even after decades of research, several aspects remain unclear, including the precise value of the localization length critical exponent \cite{Evers2008, Slevin2009, Zhu2018}, the agreement between single-particle numerics and experimental measurements \cite{Li2005, Li2009, Lutken2014}, and the nature of the effective field theory at the critical point \cite{Obuse2010, Gruzberg2017}.

The last decade has also seen burgeoning interest in many-body localization (MBL)\cite{Basko2006, Oganesyan2007, Pal2010, Nandkishore2015, Abanin2017, Imbrie2017, Alet2018, Parameswaran2018}, a generalization of Anderson localization to highly excited eigenstates of interacting many-body systems. 
However, the two fields have remained largely separated.
The presence of extended single-particle states in the Landau level has been argued to delocalize the entire many-body spectrum in the presence of interactions\cite{Nandkishore2014A}:
the topological extended states indirectly couple states localized arbitrarily far away from one other, and thus induce level repulsion across the spectrum.
Numerical exact diagonalization results for electrons in the lowest Landau level (LLL)\cite{Geraedts2017} are consistent with this prediction, pointing to the absence of an MBL phase in the thermodynamic limit. 
On the other hand, this may be due features of the LLL \emph{other} than its topological character,
{\it e.g.}, its dimensionality.

Whether or not MBL can exist in two dimensions is still an open question.
As a true thermodynamic phase, it has been argued\cite{DeRoeck2017} to be unstable towards the proliferation of rare thermal regions in dimension $d>1$ (though the issue is not settled\cite{Potirniche2018}). Nonetheless, there is experimental evidence of slow thermalization or ``glassiness'' in finite-sized two-dimensional systems\cite{Choi2016, Bordia2017}.
It is thus interesting to explore the localization properties of interacting electrons in the LLL, and clarify whether the rapid drift of the critical disorder strength observed in Ref.~\cite{Geraedts2017} is purely a consequence of the topological character, or whether it is caused by any other properties of the system ({\it e.g.} its dimensionality, or the continuum nature of the LLL orbitals).

In this work, we isolate the effect of topological states on localization, and confirm that there is  in principle no obstruction to MBL in non-topological LLL subbands (barring issues with rare-region effects in two dimensions).
These bands therefore constitute a novel setting for the study of MBL, where a finite Hilbert space is obtained through Landau level quantization in a continuum, rather than by a tight-binding discretization.

Topologically trivial subbands of the LLL can be engineered in several ways.
Periodic potentials in the LLL give rise to Hofstadter subbands with various distinct Chern numbers\cite{Hofstadter1976, Thouless1982}.
The value $C=0$ is quite commonly obtained, and by appropriately engineering the periodic potential one can ensure that the Hofstadter subband structure features a $C=0$ band which is nearly flat and well separated from the rest of the LLL, forming an ideal setting for studies of localization.

Another possibility is to lift one state at a time by inserting a point impurity (modeled by a $\delta$ function potential) in the system\cite{Ippoliti2018}. Each impurity lifts a topologically trivial bound state.
Upon inserting $N_\delta$ impurities (with $N_\delta<N_\phi$, number of fluxes through the system), 
one has a degenerate manifold of states that avoid all impurities, carrying total $C=1$, and an orthogonal manifold of $C=0$ states.
If the impurities have similar strength and are sufficiently dilute, the $C=0$ band is quite flat, and it is sensible to consider the localization problem projected to such states.

In this paper, we study the issue of MBL within these bands derived from the LLL.
The details of single particle localization in both scenarios will be discussed in an upcoming paper\cite{Krishna2018}.



\textit{Method.} 
We consider electrons in the LLL on a torus geometry with periodic boundary conditions
\begin{align}
H = \Psb \PLLL \left[ V_{\text{1-body}} + V_{\text{int}}\right] \PLLL \Psb,
\label{eq:mbham}
\end{align}
where $V_{\text{1-body}}$ and $V_{\text{int}}$ are the single-particle and interaction terms in the Hamiltonian.
$\PLLL$ is the projector onto the LLL and $\Psb$ further projects onto the eigenstates of one specified subband of the single-particle Hamiltonian.
We choose $V_{\text{int}}$ to be a Haldane $V_1$ pseudopotential interaction,
$V_{\rm int}(\mb q) = U L_1(q^2\ell_B^2) e^{-q^2\ell_B^2/2}$, where $U$ is the interaction strength, $L_1(x)$ is the Laguerre polynomial, and $\ell_B = \sqrt{eB/\hbar}$ is the magnetic length.
We study two different models for the single-particle term $V_{\text{1-body}}$.

\begin{figure}[h!]
\centering
\includegraphics[trim={0cm 0.0cm 0cm 0.0cm}, clip, width=\columnwidth]{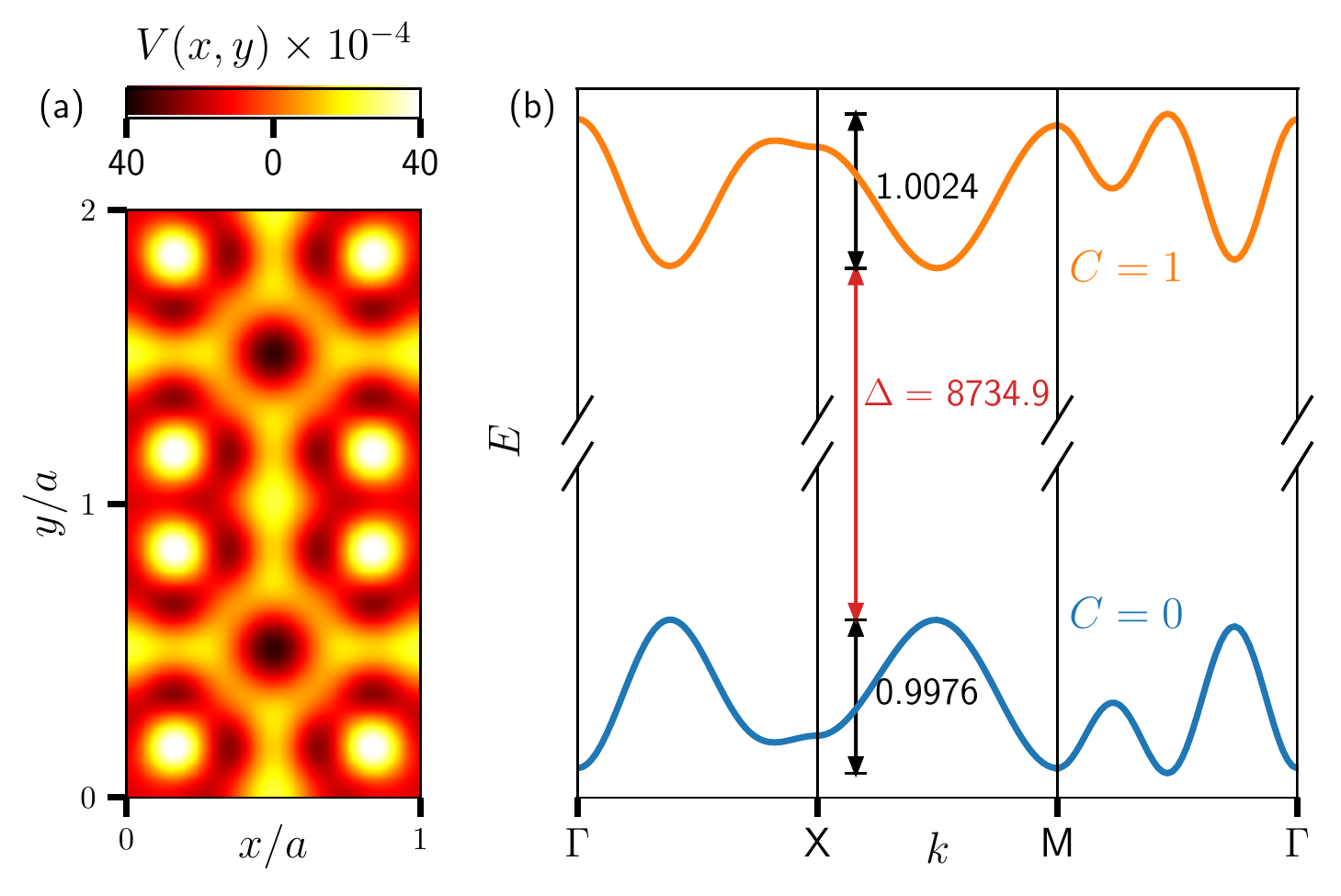}
\caption{(a) The periodic real space potential (with coefficients given in Tab.\ \ref{tab:coefficients}) is shown here over a $1 \times 2$ plaquette. (b) With $2$ magnetic flux quanta per unit cell, it leads to two nearly flat bands with Chern numbers $C=0$ and $1$ respectively.}
\label{fig:potential}
\end{figure}

First we consider the sum of a periodic potential and quenched disorder in real space,
\begin{equation}
V_{\text{1-body}}(\mb{r}) = \sum_{n_x, n_y} V_{n_x,n_y} e^{2\pi i (n_x x+n_yy)/a} 
+ V_{\rm dis} (\mb{r}) \;.
\label{eq:potential}
\end{equation}
The unit cell $a$ of the periodic potential contains two quanta of magnetic flux, i.e.\ $a^2 = 4\pi \ell_B^2$.
Such a potential splits the LLL in two subbands.
We optimize their flatness (bandwidth over bandgap ratio) in the space of coefficients $V_{n_x,n_y}$ that preserve square symmetry (keeping $n_i\leq3$),
and find optimal results at the values shown in Tab.~\ref{tab:coefficients}.
For these values of $V_{n_x,n_y}$, the potential in \eq{potential}, shown in real space in Fig.~\ref{fig:potential}(a), gives rise to two nearly flat bands.
In units of the mean bandwidth, the bandgap is $\Delta \simeq 8735$ (see Fig.~\ref{fig:potential}(b)), while the bandwidths are 1.0024 and 0.9975 respectively.
The other term in \eq{potential} is a stochastic short range correlated disorder potential, $\langle V_{\rm dis}(\mb r) V_{\rm dis}(0) \rangle = W^2 \sigma^{-2} e^{-r^2/2\sigma^2}$, where $\sigma$ is a correlation length and $W$ quantifies the strength of disorder.
We use $\sigma = 2 l_B$\footnote{ One recovers uncorrelated Gaussian white noise for $\sigma \to 0$. Setting $\sigma = 2 \ell_B$ improves the behavior of the single-particle localization length compared to $\sigma=0$ (see \cite{Krishna2018} for details)}.

\begin{table}[h!]
\begin{tabular}{c|rrrr}
 $V_{n_xn_y} \times 10^{-4}$ & $n_x=0$ & $n_x=1$ & $n_x=2$ & $n_x=3$ \\ \hline
 $n_y=0$ & $0.0000$ & $0.4156$ & $0.0000$ & $-3.5801$ \\
 $n_y=1$ & & $-0.5108$  & $0.8079$ & $-3.8980$ \\
 $n_y=2$ & & & $-0.0002$ & $2.0097$ \\
 $n_y=3$ & & & & $2.0665$
\end{tabular}
\caption{Values of potential Fourier components $V_{n_xn_y}$, from \eq{potential}, that optimize the band flatness (width-to-gap ratio) at flux per unit cell $\phi = 2$. We only show the upper triangle $n_x\geq n_y$; entries with $n_y>n_x$ are related by symmetry.
\label{tab:coefficients}}
\end{table}

\begin{figure}[h!]
\centering
\includegraphics[width=0.8\columnwidth, trim={0cm 0cm 0cm 0cm}, clip]{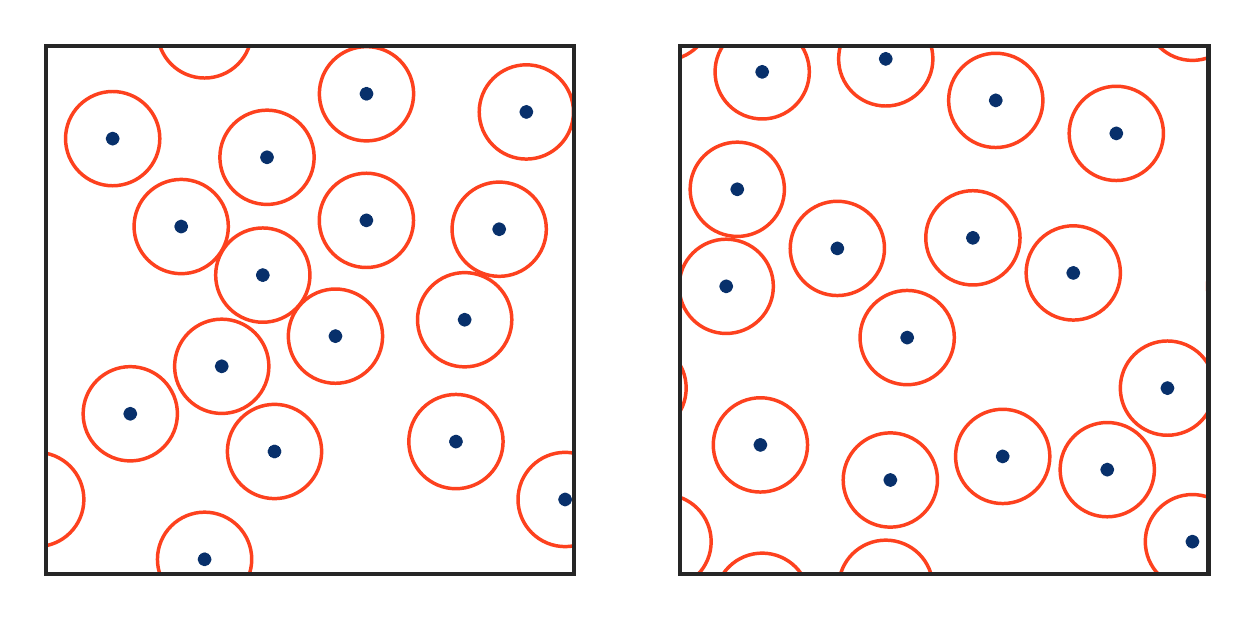}
\caption{
Two examples of configurations of $N_\delta = 16$ $\delta$-function scatterers (dark dots) randomly placed on a torus with $N_\phi = 6N_\delta$ and exclusion distance $d = 4.38\ell_B$ (red circles have radius $d/2$).
\label{fig:randomsetup}}
\end{figure}

Next, we consider a single particle Hamiltonian consisting of a number $N_\delta$ of $\delta$-function potential scatterers randomly distributed on the torus,
\begin{equation}
V_{\text{1-body}}(\mb{r}) = - \sum_{j=1}^{N_\delta} V_{j} \ \delta(\mb{r} - \mb{r}_j).
\label{eq:randomdelta}
\end{equation}
Each scatterer picks a topologically trivial state out of the Landau level.
Provided they are sufficiently dilute ($N_\delta/N_\phi$ is small enough), the $C=0$ band so obtained can be flattened to high accuracy.
Localized single-particle orbitals are obtained as a result of disorder in the positions $\mb r_j$ and amplitude $V_j$ of the scatterers.
We enforce a minimum distance $d=4.38\ell_B$ between any two scatterers, as shown in Fig.~\ref{fig:randomsetup}.
The excluded distance is chosen to be large enough that it limits rare fluctuation effects (band tails) due to clusters present in the purely Poissonian distribution, while it is small enough to allow sufficient randomness and sample to sample variations. Fig.~\ref{fig:randomsetup} shows two typical configurations for the excluded distance we choose for our current study\footnote{In hard disk terms, our ensemble corresponds to a density of 0.4, much smaller than the close-packed density for hard disks in two dimensions (0.907).}.
We randomly draw $V_j$ from a box distribution in $[1-W, 1+W]$.
$W$ and $d$ act as two independent parameters characterizing disorder.
Details of the localization problem in this two-parameter disorder space are explored further in Ref.~\cite{Krishna2018}.
For the present purpose, we note that while each realization of the spatial distribution of scatterers breaks translational and rotational symmetry, the ensemble is homogeneous and isotropic, leading to an unbiased two-dimensional scaling analysis.
The main advantage of this potential compared to the periodic one is the greater flexibility in choosing system sizes, owing to the lack of a lattice structure. Additionally, disorder here does \emph{not} mix the two subbands, so the only contribution to the projection error comes from the interaction.
The disadvantages are a worse flatness ratio, and the absence of a nearly-identical $C=1$ counterpart to the $C=0$ subband.

In either case, we set $\Psb$ in \eq{mbham} to be the projector on the subband of interest and diagonalize the projected many-body Hamiltonian.
This is the result of first-order degenerate perturbation theory\footnote{The degenerate subspace is the set of Slater determinants of orbitals in $\Psb$.} in $U/\Delta$ and $W/\Delta$, and is justified as long as $U$, $W \ll \Delta$ (bandgap). 
Contributions involving single-particle orbitals outside $\Psb$ (and hence, potentially, the topological extended states) have amplitude $\mathcal O(\Delta^{-1})$, and induce non-local Hamiltonian matrix elements at order $\mathcal O(\Delta^{-2})$, which are discarded by the projection. 

We obtain the full eigenvalue spectrum of the projected many-body Hamiltonian via exact diagonalization, and compute the average level spacing ratio $\langle r\rangle$, where $r_n = \frac{\text{min}(\delta E_n, \delta E_{n+1})}{\text{max}(\delta E_n, \delta E_{n+1})}$ and $\delta E_n = E_{n+1} - E_n$.
This quantity is a diagnostic of level repulsion \cite{Atas2013, Oganesyan2007, Pal2010, Geraedts2017}:
a localized system is well described by the Poisson random matrix ensemble, for which $\langle r \rangle \simeq 0.386$;
an ergodic system on the other hand shows level repulsion and has $\langle r \rangle \simeq 0.600$ (Gaussian unitary ensemble, GUE) in the absence of time reversal symmetry, which is the case of interest here.
We generate an ensemble of quenched disorder realizations;
for each realization we average $r_n$ over the central $1/6$ of the integrated density of states, and finally average the result over the disorder ensemble.
The \emph{sample-to-sample} standard deviation is used as a measure of the uncertainty.

\begin{figure}[h!]
\centering
\includegraphics[trim={0 0.9cm 0 0.6cm}, clip, width=\columnwidth]{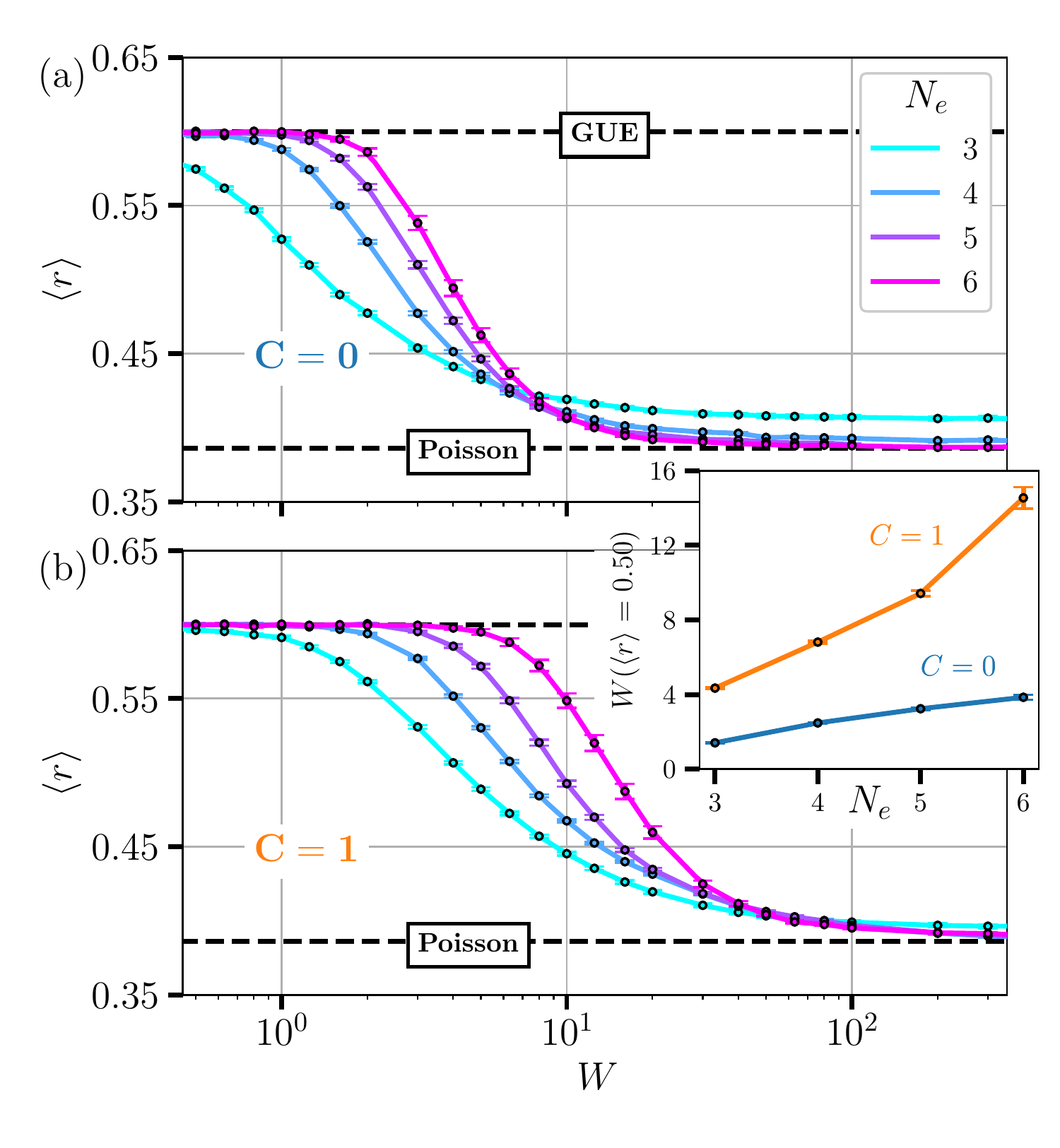}
\caption{The mean $\langle r \rangle$ statistic as a function of $W$ at filling $\nu = 1/3$ of the (a) $C = 0$ and (b) $C = 1$ band. The torus is square at $N_e=3$; for $N_e>3$, one side is kept fixed while the other is scaled, approaching a 1D thermodynamic limit.
The inset shows the dependence of critical disorder $W(\langle r\rangle=0.5)$ on system size.
\label{fig:1d}}
\end{figure}

\textit{Results.}
For the periodic potential, \eq{potential}, we fix the interaction strength $U = 8$, so that it is larger than the bandwidth, but much smaller than the bandgap of the single-particle Hamiltonian.
We place the system on a rectangular torus and increase one of the sides while leaving the other fixed.
In the thermodynamic limit, this becomes an infinite cylinder with fixed circumference.
As a one-dimensional problem with short-range interactions and quenched disorder,
when projected into a non-topological $C=0$ band, this is expected to exhibit an ergodic-to-MBL transition at a finite value of disorder strength $W_c$.
We consider rectangular tori with $3\times n$ unit cells of the potential, where $n=3,4,5,6$;
this gives a total of $N_\phi = 6n$ fluxes through the torus and $N_\phi/2 = 3n$ single-particle states in each band. 
We consider $N_e = n$ electrons (i.e., filling of the band is $\nu = 1/3$) and project onto the $C=0$ band.
Results for $\langle r \rangle$ are shown in Fig.~\ref{fig:1d}(a) and clearly display a transition at a finite critical disorder $W_c \approx 8$.
Moreover the value of $W$ at which $\langle r \rangle = 0.5$ (roughly halfway between Poisson and GUE) shows signs of saturating to a finite value as $N_e \to \infty$.
On the contrary, when projecting onto the $C=1$ band, the situation is radically different.
The data, shown in Fig.~\ref{fig:1d}(b), features no crossing and a rapid drift of the crossover disorder strength $W(\langle r \rangle = 0.5)$ to infinity with system size.
This is consistent with existing results for the entire LLL\cite{Geraedts2017}, and in line with arguments suggesting that a divergence in the single particle localization length is enough to delocalize the entire many-body spectrum.

Because of lattice commensurability constraints and limits on system sizes amenable to exact diagonalization, a two-dimensional scaling study for this model is not able to yield conclusive results at present.
The C = 1 band is clearly found to delocalize; however, the fate of the C = 0 band for this model remains unclear with the sizes we have access to at this stage.

To circumvent this limitation, we turn to the $C=0$ band obtained from random $\delta$ scatterers, \eq{randomdelta}.
This allows us to perform two-dimensional scaling with more flexibility in the choice of sizes.
We take one $\delta$-function scatterer per 6 quanta of magnetic flux ($N_\phi = 6N_\delta$) and set the electron filling to $\nu = 1/2$ ($N_\delta = 2N_e$).
The minimum distance between $\delta$-function scatterers is set to $d=4.38\ell_B$.
These parameters are found to give a single-particle $C=0$ band with average flatness ratio $f\simeq 10^{-2}$ when $W=0$ (\textit{i.e.} when all scatterers have the same strength).

\begin{figure}
\centering
\includegraphics[width=\columnwidth, trim={0 0 0 0}, clip]{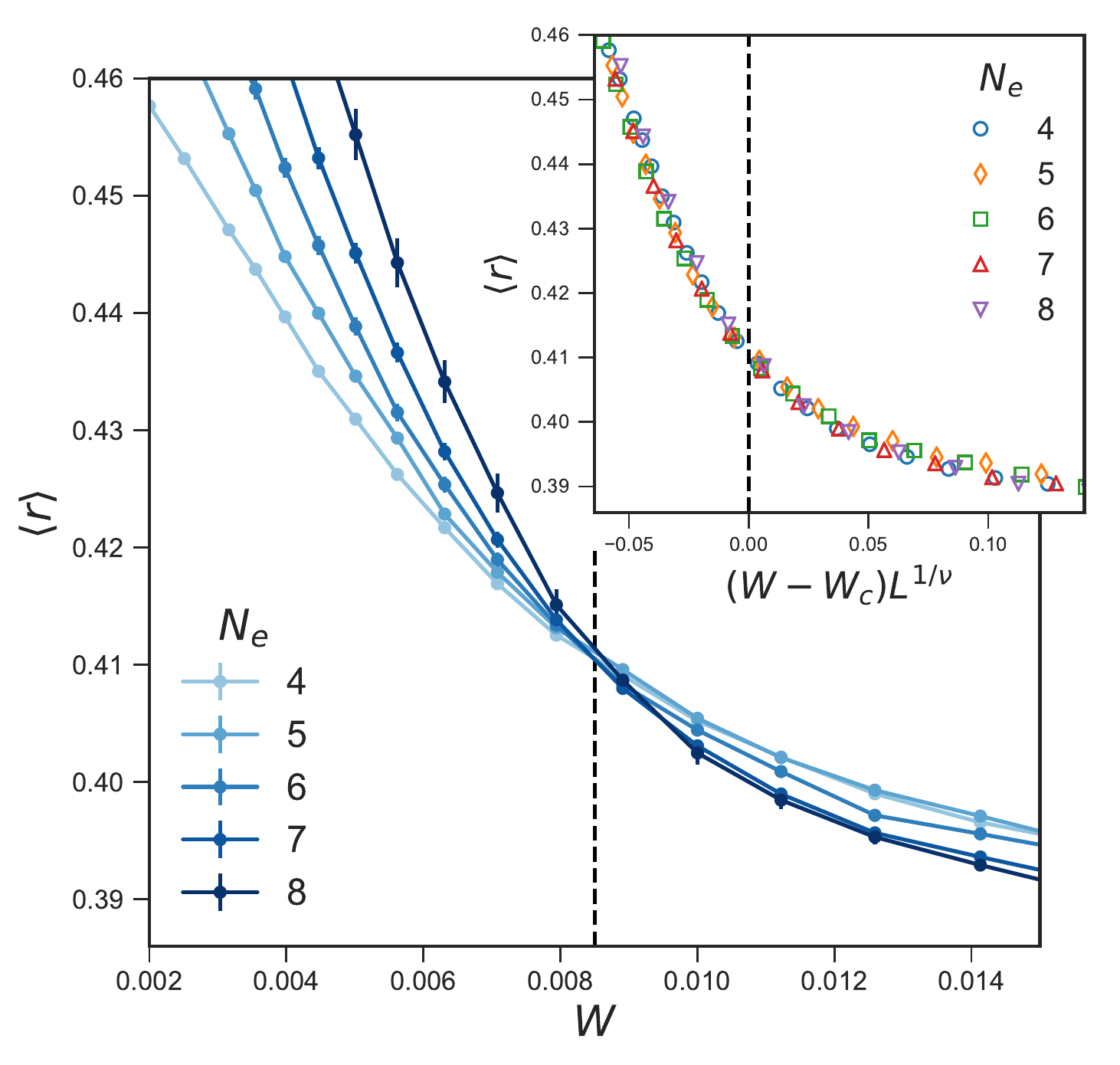}
\caption{
Level spacing ratio $\langle r \rangle$ as a function of disorder $W$ for the random scatterer model of \eq{randomdelta}.
$N_e$ electrons are projected in the topologically trivial band created by $2N_e$ $\delta$-function scatterers (filling is $\nu = 1/2$) on a rectangular torus pierced by $12N_e$ magnetic fluxes. The minimum distance between two scatterers is fixed at $d = 4.38 \ell_B$. 
At $N_e = 4$ the torus is square, then one side is kept fixed as the other is increased.
The data for different sizes cross at $W_c \simeq 8.5\times 10^{-3}$.
Inset: scaling collapse of the data in the main plot. The localization length exponent $\nu$ that achieves the best collapse is $\nu = 1.3 \pm 0.1$.
\label{fig:random}}
\end{figure}

\begin{figure}
\centering
\includegraphics[width=\columnwidth, trim={0 0 0 0}, clip]{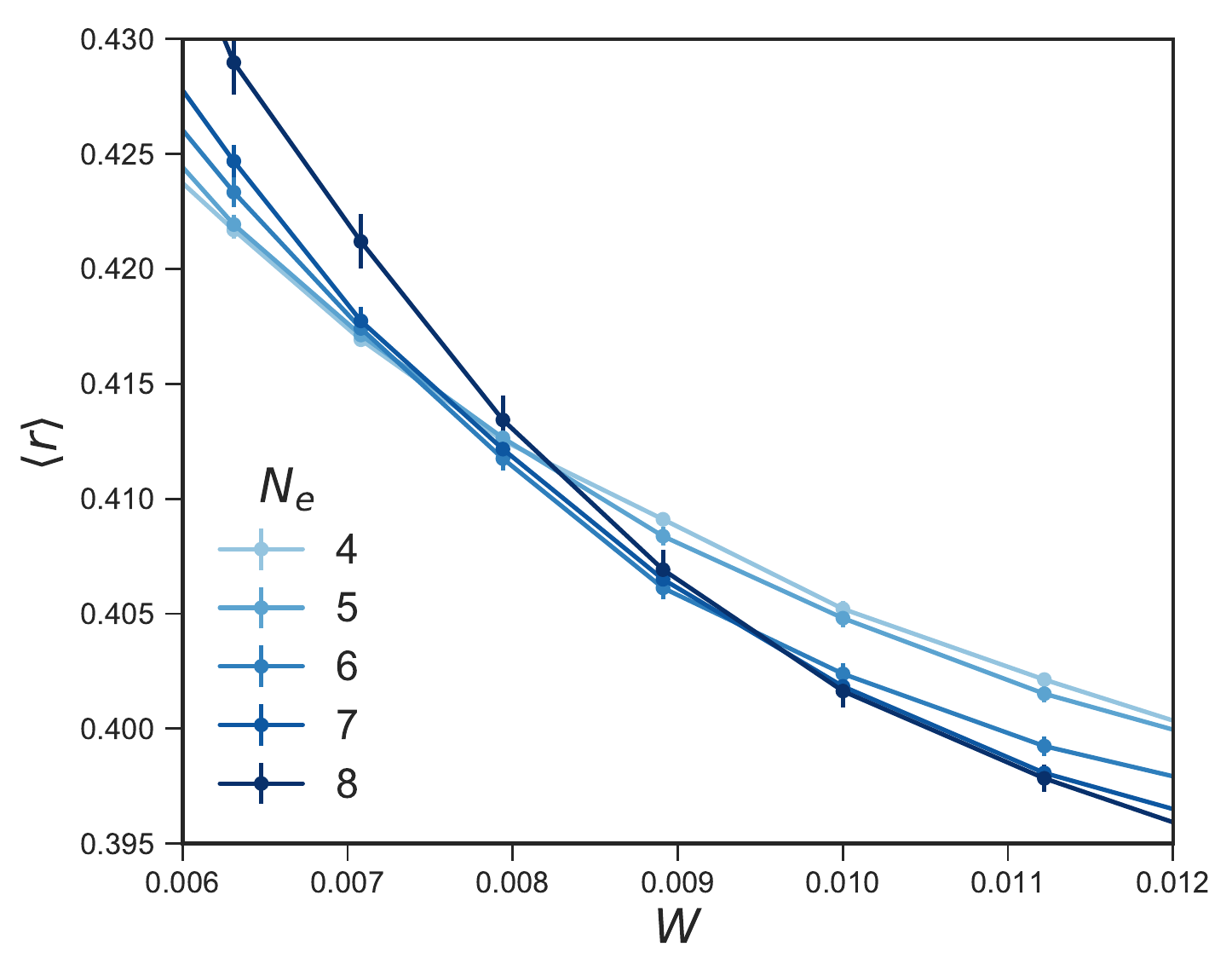}
\caption{
Level spacing ratio $\langle r \rangle$ as a function of disorder $W$ for the same model as Fig.~\ref{fig:random}, but for two dimensional scaling of square tori.
Curves for consecutive sizes cross at values of $W$ that drift to the right, from $W \simeq 7.5\times 10^{-3}$ ($N_e = 4$ and 5) to $W_c\simeq 1.0 \times 10^{-2}$ ($N_e=7$ and 8).
\label{fig:random2d}}
\end{figure}

We set $N_e = 4$ and consider a square torus. 
From there, we perform both one-dimensional and two-dimensional finite-size scaling:
in one case we increase only one side of the torus, making it rectangular, Fig.~\ref{fig:random};
in the other we maintain the square aspect ratio, Fig.~\ref{fig:random2d}.
We go up to $N_e = 8$ and we average $5\times 10^4$ realizations at the smallest size and $10^3$ at the largest (in each realization the scatterers have independent positions \emph{and} strengths).
We see clear indications of a transition at a finite disorder strength of $W_c \simeq 0.0085$ in the 1D case.
A single-parameter scaling collapse yields the critical exponent $\nu_{1D} \approx 1.3$, in violation of the CCFS criterion\cite{Chayes1986, Chandran2015} $\nu \geq 2/d$ for disorder-driven transitions. 
This violation seems to be common in exact diagonalization studies of MBL in one-dimensional spin chains \cite{Kjall2014, Luitz2015}.
The relationship between these results and the CCFS bound, as well as strong-disorder renormalization group studies suggesting a much higher value of $\nu$\cite{Vosk2015, Potter2015, Dumitrescu2017}, is unclear.
Moreover, here as well as in the periodic potential case of Fig.~\ref{fig:potential}, we find $\langle r\rangle \simeq 0.41$ at the transition, very close to the Poisson value and in line with previous findings in spin systems\cite{Khemani2017}, suggesting the value may be universal at the MBL transition.

The two-dimensional scaling yields similar data with a clear inversion in the order of sizes beyond $W\approx 0.01$, but the crossings of consecutive sizes slowly drift towards larger $W$ (see Fig.~\ref{fig:random2d}). 
If the drift slows down at larger sizes, this may simply signal that the transition in the thermodynamic limit is happening at larger $W$ compared to the one dimensional case, which would be expected.
However, it is also possible that the drift may continue indefinitely and signal instability of the MBL phase in two dimensions. 
Results on larger sizes are needed to establish conclusively if there is lack of ergodicity in the two-dimensional thermodynamic limit.


\textit{Conclusion}.
We have investigated the problem of many-body localization of electrons in the lowest Landau level in the absence of single-particle extended states.
We have removed such states, which are known to delocalize the many-body spectrum, by isolating a nearly-flat, topologically trivial ($C=0$) band in two different ways:
(i) by using a suitably engineered periodic potential, and 
(ii) by means of dilute, randomly distributed $\delta$-function (point-like) scatterers. 
Both methods yield strong evidence of a transition in the one-dimensional (cylindrical) thermodynamic limit, showing non-topological Landau level subbands as a novel, \emph{continuum} seeting for many-body localization.
Moreover, method (ii) points to the same scenario for the two-dimensional (planar) thermodynamic limit. 
It remains to be clarified whether the transition point in the latter case is drifting towards finite or infinite disorder with increasing size.

It should be noted that the Landau level subbands that we project out generally introduce non-local Hamiltonian matrix elements with strength $\sim E^3\Delta^{-2}$, where $\Delta$ is the bandgap 
and $E$ is the largest energy scale in the projected problem.
These non-local couplings are expected to drive thermalization over a time scale $t \sim \hbar \Delta^2/E^3$, which is long but finite. Thus ``localization'' in the present context should be more appropriately understood as slow thermalization.
Nonetheless, $\Delta/E$ can in principle be made arbitrarily large, leading to an arbitrarily long transient in which the system is effectively localized.

Finally, in this work we have used the level spacing ratio $r$ to diagnose lack of ergodicity. 
It would be interesting to test dynamical features of the MBL phase, such as memory of initial conditions, logarithmic spreading of entanglement, and emergent integrability, in the setting we have proposed. We leave these problems for future work.

\textit{Acknowledgments.} 
This work was supported by DOE BES grant $\text{DE-SC0002140}$.
We acknowledge useful conversations with Rahul Nandkishore.

\bibliography{qh_loc}

\begin{thebibliography}{42}%
\makeatletter
\providecommand \@ifxundefined [1]{%
 \@ifx{#1\undefined}
}%
\providecommand \@ifnum [1]{%
 \ifnum #1\expandafter \@firstoftwo
 \else \expandafter \@secondoftwo
 \fi
}%
\providecommand \@ifx [1]{%
 \ifx #1\expandafter \@firstoftwo
 \else \expandafter \@secondoftwo
 \fi
}%
\providecommand \natexlab [1]{#1}%
\providecommand \enquote  [1]{``#1''}%
\providecommand \bibnamefont  [1]{#1}%
\providecommand \bibfnamefont [1]{#1}%
\providecommand \citenamefont [1]{#1}%
\providecommand \href@noop [0]{\@secondoftwo}%
\providecommand \href [0]{\begingroup \@sanitize@url \@href}%
\providecommand \@href[1]{\@@startlink{#1}\@@href}%
\providecommand \@@href[1]{\endgroup#1\@@endlink}%
\providecommand \@sanitize@url [0]{\catcode `\\12\catcode `\$12\catcode
  `\&12\catcode `\#12\catcode `\^12\catcode `\_12\catcode `\%12\relax}%
\providecommand \@@startlink[1]{}%
\providecommand \@@endlink[0]{}%
\providecommand \url  [0]{\begingroup\@sanitize@url \@url }%
\providecommand \@url [1]{\endgroup\@href {#1}{\urlprefix }}%
\providecommand \urlprefix  [0]{URL }%
\providecommand \Eprint [0]{\href }%
\providecommand \doibase [0]{http://dx.doi.org/}%
\providecommand \selectlanguage [0]{\@gobble}%
\providecommand \bibinfo  [0]{\@secondoftwo}%
\providecommand \bibfield  [0]{\@secondoftwo}%
\providecommand \translation [1]{[#1]}%
\providecommand \BibitemOpen [0]{}%
\providecommand \bibitemStop [0]{}%
\providecommand \bibitemNoStop [0]{.\EOS\space}%
\providecommand \EOS [0]{\spacefactor3000\relax}%
\providecommand \BibitemShut  [1]{\csname bibitem#1\endcsname}%
\let\auto@bib@innerbib\@empty
\bibitem [{\citenamefont {Prange}(1981)}]{Prange1981}%
  \BibitemOpen
  \bibfield  {author} {\bibinfo {author} {\bibfnamefont {R.~E.}\ \bibnamefont
  {Prange}},\ }\href {\doibase 10.1103/PhysRevB.23.4802} {\bibfield  {journal}
  {\bibinfo  {journal} {Phys. Rev. B}\ }\textbf {\bibinfo {volume} {23}},\
  \bibinfo {pages} {4802} (\bibinfo {year} {1981})}\BibitemShut {NoStop}%
\bibitem [{\citenamefont {Chalker}\ and\ \citenamefont
  {Coddington}(1988)}]{Chalker1988}%
  \BibitemOpen
  \bibfield  {author} {\bibinfo {author} {\bibfnamefont {J.~T.}\ \bibnamefont
  {Chalker}}\ and\ \bibinfo {author} {\bibfnamefont {P.~D.}\ \bibnamefont
  {Coddington}},\ }\href {http://stacks.iop.org/0022-3719/21/i=14/a=008}
  {\bibfield  {journal} {\bibinfo  {journal} {Journal of Physics C: Solid State
  Physics}\ }\textbf {\bibinfo {volume} {21}},\ \bibinfo {pages} {2665}
  (\bibinfo {year} {1988})}\BibitemShut {NoStop}%
\bibitem [{\citenamefont {Huo}\ and\ \citenamefont {Bhatt}(1992)}]{Huo1992}%
  \BibitemOpen
  \bibfield  {author} {\bibinfo {author} {\bibfnamefont {Y.}~\bibnamefont
  {Huo}}\ and\ \bibinfo {author} {\bibfnamefont {R.~N.}\ \bibnamefont
  {Bhatt}},\ }\href {\doibase 10.1103/PhysRevLett.68.1375} {\bibfield
  {journal} {\bibinfo  {journal} {Phys. Rev. Lett.}\ }\textbf {\bibinfo
  {volume} {68}},\ \bibinfo {pages} {1375} (\bibinfo {year}
  {1992})}\BibitemShut {NoStop}%
\bibitem [{\citenamefont {Huckestein}(1995)}]{Huckestein1995}%
  \BibitemOpen
  \bibfield  {author} {\bibinfo {author} {\bibfnamefont {B.}~\bibnamefont
  {Huckestein}},\ }\href {\doibase 10.1103/RevModPhys.67.357} {\bibfield
  {journal} {\bibinfo  {journal} {Rev. Mod. Phys.}\ }\textbf {\bibinfo {volume}
  {67}},\ \bibinfo {pages} {357} (\bibinfo {year} {1995})}\BibitemShut
  {NoStop}%
\bibitem [{\citenamefont {Evers}\ and\ \citenamefont
  {Mirlin}(2008)}]{Evers2008}%
  \BibitemOpen
  \bibfield  {author} {\bibinfo {author} {\bibfnamefont {F.}~\bibnamefont
  {Evers}}\ and\ \bibinfo {author} {\bibfnamefont {A.~D.}\ \bibnamefont
  {Mirlin}},\ }\href {\doibase 10.1103/RevModPhys.80.1355} {\bibfield
  {journal} {\bibinfo  {journal} {Rev. Mod. Phys.}\ }\textbf {\bibinfo {volume}
  {80}},\ \bibinfo {pages} {1355} (\bibinfo {year} {2008})}\BibitemShut
  {NoStop}%
\bibitem [{\citenamefont {Slevin}\ and\ \citenamefont
  {Ohtsuki}(2009)}]{Slevin2009}%
  \BibitemOpen
  \bibfield  {author} {\bibinfo {author} {\bibfnamefont {K.}~\bibnamefont
  {Slevin}}\ and\ \bibinfo {author} {\bibfnamefont {T.}~\bibnamefont
  {Ohtsuki}},\ }\href {\doibase 10.1103/PhysRevB.80.041304} {\bibfield
  {journal} {\bibinfo  {journal} {Phys. Rev. B}\ }\textbf {\bibinfo {volume}
  {80}},\ \bibinfo {pages} {041304} (\bibinfo {year} {2009})}\BibitemShut
  {NoStop}%
\bibitem [{\citenamefont {{Zhu}}\ \emph {et~al.}(2018)\citenamefont {{Zhu}},
  \citenamefont {{Wu}}, \citenamefont {{Bhatt}},\ and\ \citenamefont
  {{Wan}}}]{Zhu2018}%
  \BibitemOpen
  \bibfield  {author} {\bibinfo {author} {\bibfnamefont {Q.}~\bibnamefont
  {{Zhu}}}, \bibinfo {author} {\bibfnamefont {P.}~\bibnamefont {{Wu}}},
  \bibinfo {author} {\bibfnamefont {R.~N.}\ \bibnamefont {{Bhatt}}}, \ and\
  \bibinfo {author} {\bibfnamefont {X.}~\bibnamefont {{Wan}}},\ }\href@noop {}
  {\bibfield  {journal} {\bibinfo  {journal} {ArXiv e-prints}\ } (\bibinfo
  {year} {2018})},\ \Eprint {http://arxiv.org/abs/1804.00398} {arXiv:1804.00398
  [cond-mat.dis-nn]} \BibitemShut {NoStop}%
\bibitem [{\citenamefont {Li}\ \emph {et~al.}(2005)\citenamefont {Li},
  \citenamefont {Cs\'athy}, \citenamefont {Tsui}, \citenamefont {Pfeiffer},\
  and\ \citenamefont {West}}]{Li2005}%
  \BibitemOpen
  \bibfield  {author} {\bibinfo {author} {\bibfnamefont {W.}~\bibnamefont
  {Li}}, \bibinfo {author} {\bibfnamefont {G.~A.}\ \bibnamefont {Cs\'athy}},
  \bibinfo {author} {\bibfnamefont {D.~C.}\ \bibnamefont {Tsui}}, \bibinfo
  {author} {\bibfnamefont {L.~N.}\ \bibnamefont {Pfeiffer}}, \ and\ \bibinfo
  {author} {\bibfnamefont {K.~W.}\ \bibnamefont {West}},\ }\href {\doibase
  10.1103/PhysRevLett.94.206807} {\bibfield  {journal} {\bibinfo  {journal}
  {Phys. Rev. Lett.}\ }\textbf {\bibinfo {volume} {94}},\ \bibinfo {pages}
  {206807} (\bibinfo {year} {2005})}\BibitemShut {NoStop}%
\bibitem [{\citenamefont {Li}\ \emph {et~al.}(2009)\citenamefont {Li},
  \citenamefont {Vicente}, \citenamefont {Xia}, \citenamefont {Pan},
  \citenamefont {Tsui}, \citenamefont {Pfeiffer},\ and\ \citenamefont
  {West}}]{Li2009}%
  \BibitemOpen
  \bibfield  {author} {\bibinfo {author} {\bibfnamefont {W.}~\bibnamefont
  {Li}}, \bibinfo {author} {\bibfnamefont {C.~L.}\ \bibnamefont {Vicente}},
  \bibinfo {author} {\bibfnamefont {J.~S.}\ \bibnamefont {Xia}}, \bibinfo
  {author} {\bibfnamefont {W.}~\bibnamefont {Pan}}, \bibinfo {author}
  {\bibfnamefont {D.~C.}\ \bibnamefont {Tsui}}, \bibinfo {author}
  {\bibfnamefont {L.~N.}\ \bibnamefont {Pfeiffer}}, \ and\ \bibinfo {author}
  {\bibfnamefont {K.~W.}\ \bibnamefont {West}},\ }\href {\doibase
  10.1103/PhysRevLett.102.216801} {\bibfield  {journal} {\bibinfo  {journal}
  {Phys. Rev. Lett.}\ }\textbf {\bibinfo {volume} {102}},\ \bibinfo {pages}
  {216801} (\bibinfo {year} {2009})}\BibitemShut {NoStop}%
\bibitem [{\citenamefont {Lütken}\ and\ \citenamefont
  {Ross}(2014)}]{Lutken2014}%
  \BibitemOpen
  \bibfield  {author} {\bibinfo {author} {\bibfnamefont {C.}~\bibnamefont
  {Lütken}}\ and\ \bibinfo {author} {\bibfnamefont {G.}~\bibnamefont {Ross}},\
  }\href {\doibase https://doi.org/10.1016/j.physleta.2013.11.001} {\bibfield
  {journal} {\bibinfo  {journal} {Physics Letters A}\ }\textbf {\bibinfo
  {volume} {378}},\ \bibinfo {pages} {262 } (\bibinfo {year}
  {2014})}\BibitemShut {NoStop}%
\bibitem [{\citenamefont {Obuse}\ \emph {et~al.}(2010)\citenamefont {Obuse},
  \citenamefont {Subramaniam}, \citenamefont {Furusaki}, \citenamefont
  {Gruzberg},\ and\ \citenamefont {Ludwig}}]{Obuse2010}%
  \BibitemOpen
  \bibfield  {author} {\bibinfo {author} {\bibfnamefont {H.}~\bibnamefont
  {Obuse}}, \bibinfo {author} {\bibfnamefont {A.~R.}\ \bibnamefont
  {Subramaniam}}, \bibinfo {author} {\bibfnamefont {A.}~\bibnamefont
  {Furusaki}}, \bibinfo {author} {\bibfnamefont {I.~A.}\ \bibnamefont
  {Gruzberg}}, \ and\ \bibinfo {author} {\bibfnamefont {A.~W.~W.}\ \bibnamefont
  {Ludwig}},\ }\href {\doibase 10.1103/PhysRevB.82.035309} {\bibfield
  {journal} {\bibinfo  {journal} {Phys. Rev. B}\ }\textbf {\bibinfo {volume}
  {82}},\ \bibinfo {pages} {035309} (\bibinfo {year} {2010})}\BibitemShut
  {NoStop}%
\bibitem [{\citenamefont {Gruzberg}\ \emph {et~al.}(2017)\citenamefont
  {Gruzberg}, \citenamefont {Kl\"umper}, \citenamefont {Nuding},\ and\
  \citenamefont {Sedrakyan}}]{Gruzberg2017}%
  \BibitemOpen
  \bibfield  {author} {\bibinfo {author} {\bibfnamefont {I.~A.}\ \bibnamefont
  {Gruzberg}}, \bibinfo {author} {\bibfnamefont {A.}~\bibnamefont {Kl\"umper}},
  \bibinfo {author} {\bibfnamefont {W.}~\bibnamefont {Nuding}}, \ and\ \bibinfo
  {author} {\bibfnamefont {A.}~\bibnamefont {Sedrakyan}},\ }\href {\doibase
  10.1103/PhysRevB.95.125414} {\bibfield  {journal} {\bibinfo  {journal} {Phys.
  Rev. B}\ }\textbf {\bibinfo {volume} {95}},\ \bibinfo {pages} {125414}
  (\bibinfo {year} {2017})}\BibitemShut {NoStop}%
\bibitem [{\citenamefont {Basko}\ \emph {et~al.}(2006)\citenamefont {Basko},
  \citenamefont {Aleiner},\ and\ \citenamefont {Altshuler}}]{Basko2006}%
  \BibitemOpen
  \bibfield  {author} {\bibinfo {author} {\bibfnamefont {D.}~\bibnamefont
  {Basko}}, \bibinfo {author} {\bibfnamefont {I.}~\bibnamefont {Aleiner}}, \
  and\ \bibinfo {author} {\bibfnamefont {B.}~\bibnamefont {Altshuler}},\ }\href
  {\doibase https://doi.org/10.1016/j.aop.2005.11.014} {\bibfield  {journal}
  {\bibinfo  {journal} {Annals of Physics}\ }\textbf {\bibinfo {volume}
  {321}},\ \bibinfo {pages} {1126 } (\bibinfo {year} {2006})}\BibitemShut
  {NoStop}%
\bibitem [{\citenamefont {Oganesyan}\ and\ \citenamefont
  {Huse}(2007)}]{Oganesyan2007}%
  \BibitemOpen
  \bibfield  {author} {\bibinfo {author} {\bibfnamefont {V.}~\bibnamefont
  {Oganesyan}}\ and\ \bibinfo {author} {\bibfnamefont {D.~A.}\ \bibnamefont
  {Huse}},\ }\href {\doibase 10.1103/PhysRevB.75.155111} {\bibfield  {journal}
  {\bibinfo  {journal} {Phys. Rev. B}\ }\textbf {\bibinfo {volume} {75}},\
  \bibinfo {pages} {155111} (\bibinfo {year} {2007})}\BibitemShut {NoStop}%
\bibitem [{\citenamefont {Pal}\ and\ \citenamefont {Huse}(2010)}]{Pal2010}%
  \BibitemOpen
  \bibfield  {author} {\bibinfo {author} {\bibfnamefont {A.}~\bibnamefont
  {Pal}}\ and\ \bibinfo {author} {\bibfnamefont {D.~A.}\ \bibnamefont {Huse}},\
  }\href {\doibase 10.1103/PhysRevB.82.174411} {\bibfield  {journal} {\bibinfo
  {journal} {Phys. Rev. B}\ }\textbf {\bibinfo {volume} {82}},\ \bibinfo
  {pages} {174411} (\bibinfo {year} {2010})}\BibitemShut {NoStop}%
\bibitem [{\citenamefont {Nandkishore}\ and\ \citenamefont
  {Huse}(2015)}]{Nandkishore2015}%
  \BibitemOpen
  \bibfield  {author} {\bibinfo {author} {\bibfnamefont {R.}~\bibnamefont
  {Nandkishore}}\ and\ \bibinfo {author} {\bibfnamefont {D.~A.}\ \bibnamefont
  {Huse}},\ }\href {\doibase 10.1146/annurev-conmatphys-031214-014726}
  {\bibfield  {journal} {\bibinfo  {journal} {Annual Review of Condensed Matter
  Physics}\ }\textbf {\bibinfo {volume} {6}},\ \bibinfo {pages} {15} (\bibinfo
  {year} {2015})}\BibitemShut {NoStop}%
\bibitem [{\citenamefont {Abanin}\ and\ \citenamefont
  {Papić}(2017)}]{Abanin2017}%
  \BibitemOpen
  \bibfield  {author} {\bibinfo {author} {\bibfnamefont {D.~A.}\ \bibnamefont
  {Abanin}}\ and\ \bibinfo {author} {\bibfnamefont {Z.}~\bibnamefont
  {Papić}},\ }\href {\doibase 10.1002/andp.201700169} {\bibfield  {journal}
  {\bibinfo  {journal} {Annalen der Physik}\ }\textbf {\bibinfo {volume}
  {529}},\ \bibinfo {pages} {1700169} (\bibinfo {year} {2017})}\BibitemShut
  {NoStop}%
\bibitem [{\citenamefont {Imbrie}\ \emph {et~al.}(2017)\citenamefont {Imbrie},
  \citenamefont {Ros},\ and\ \citenamefont {Scardicchio}}]{Imbrie2017}%
  \BibitemOpen
  \bibfield  {author} {\bibinfo {author} {\bibfnamefont {J.~Z.}\ \bibnamefont
  {Imbrie}}, \bibinfo {author} {\bibfnamefont {V.}~\bibnamefont {Ros}}, \ and\
  \bibinfo {author} {\bibfnamefont {A.}~\bibnamefont {Scardicchio}},\ }\href
  {\doibase 10.1002/andp.201600278} {\bibfield  {journal} {\bibinfo  {journal}
  {Annalen der Physik}\ }\textbf {\bibinfo {volume} {529}},\ \bibinfo {pages}
  {1600278} (\bibinfo {year} {2017})}\BibitemShut {NoStop}%
\bibitem [{\citenamefont {Alet}\ and\ \citenamefont
  {Laflorencie}(2018)}]{Alet2018}%
  \BibitemOpen
  \bibfield  {author} {\bibinfo {author} {\bibfnamefont {F.}~\bibnamefont
  {Alet}}\ and\ \bibinfo {author} {\bibfnamefont {N.}~\bibnamefont
  {Laflorencie}},\ }\href {\doibase https://doi.org/10.1016/j.crhy.2018.03.003}
  {\bibfield  {journal} {\bibinfo  {journal} {Comptes Rendus Physique}\ }
  (\bibinfo {year} {2018}),\
  https://doi.org/10.1016/j.crhy.2018.03.003}\BibitemShut {NoStop}%
\bibitem [{\citenamefont {Parameswaran}\ and\ \citenamefont
  {Vasseur}(2018)}]{Parameswaran2018}%
  \BibitemOpen
  \bibfield  {author} {\bibinfo {author} {\bibfnamefont {S.~A.}\ \bibnamefont
  {Parameswaran}}\ and\ \bibinfo {author} {\bibfnamefont {R.}~\bibnamefont
  {Vasseur}},\ }\href {http://stacks.iop.org/0034-4885/81/i=8/a=082501}
  {\bibfield  {journal} {\bibinfo  {journal} {Reports on Progress in Physics}\
  }\textbf {\bibinfo {volume} {81}},\ \bibinfo {pages} {082501} (\bibinfo
  {year} {2018})}\BibitemShut {NoStop}%
\bibitem [{\citenamefont {Nandkishore}\ and\ \citenamefont
  {Potter}(2014)}]{Nandkishore2014A}%
  \BibitemOpen
  \bibfield  {author} {\bibinfo {author} {\bibfnamefont {R.}~\bibnamefont
  {Nandkishore}}\ and\ \bibinfo {author} {\bibfnamefont {A.~C.}\ \bibnamefont
  {Potter}},\ }\href {\doibase 10.1103/PhysRevB.90.195115} {\bibfield
  {journal} {\bibinfo  {journal} {Phys. Rev. B}\ }\textbf {\bibinfo {volume}
  {90}},\ \bibinfo {pages} {195115} (\bibinfo {year} {2014})}\BibitemShut
  {NoStop}%
\bibitem [{\citenamefont {Geraedts}\ and\ \citenamefont
  {Bhatt}(2017)}]{Geraedts2017}%
  \BibitemOpen
  \bibfield  {author} {\bibinfo {author} {\bibfnamefont {S.~D.}\ \bibnamefont
  {Geraedts}}\ and\ \bibinfo {author} {\bibfnamefont {R.~N.}\ \bibnamefont
  {Bhatt}},\ }\href {\doibase 10.1103/PhysRevB.95.054303} {\bibfield  {journal}
  {\bibinfo  {journal} {Phys. Rev. B}\ }\textbf {\bibinfo {volume} {95}},\
  \bibinfo {pages} {054303} (\bibinfo {year} {2017})}\BibitemShut {NoStop}%
\bibitem [{\citenamefont {De~Roeck}\ and\ \citenamefont
  {Huveneers}(2017)}]{DeRoeck2017}%
  \BibitemOpen
  \bibfield  {author} {\bibinfo {author} {\bibfnamefont {W.}~\bibnamefont
  {De~Roeck}}\ and\ \bibinfo {author} {\bibfnamefont {F.}~\bibnamefont
  {Huveneers}},\ }\href {\doibase 10.1103/PhysRevB.95.155129} {\bibfield
  {journal} {\bibinfo  {journal} {Phys. Rev. B}\ }\textbf {\bibinfo {volume}
  {95}},\ \bibinfo {pages} {155129} (\bibinfo {year} {2017})}\BibitemShut
  {NoStop}%
\bibitem [{\citenamefont {{Potirniche}}\ \emph {et~al.}(2018)\citenamefont
  {{Potirniche}}, \citenamefont {{Banerjee}},\ and\ \citenamefont
  {{Altman}}}]{Potirniche2018}%
  \BibitemOpen
  \bibfield  {author} {\bibinfo {author} {\bibfnamefont {I.-D.}\ \bibnamefont
  {{Potirniche}}}, \bibinfo {author} {\bibfnamefont {S.}~\bibnamefont
  {{Banerjee}}}, \ and\ \bibinfo {author} {\bibfnamefont {E.}~\bibnamefont
  {{Altman}}},\ }\href@noop {} {\bibfield  {journal} {\bibinfo  {journal}
  {ArXiv e-prints}\ } (\bibinfo {year} {2018})},\ \Eprint
  {http://arxiv.org/abs/1805.01475} {arXiv:1805.01475 [cond-mat.dis-nn]}
  \BibitemShut {NoStop}%
\bibitem [{\citenamefont {Choi}\ \emph {et~al.}(2016)\citenamefont {Choi},
  \citenamefont {Hild}, \citenamefont {Zeiher}, \citenamefont {Schau{\ss}},
  \citenamefont {Rubio-Abadal}, \citenamefont {Yefsah}, \citenamefont
  {Khemani}, \citenamefont {Huse}, \citenamefont {Bloch},\ and\ \citenamefont
  {Gross}}]{Choi2016}%
  \BibitemOpen
  \bibfield  {author} {\bibinfo {author} {\bibfnamefont {J.-y.}\ \bibnamefont
  {Choi}}, \bibinfo {author} {\bibfnamefont {S.}~\bibnamefont {Hild}}, \bibinfo
  {author} {\bibfnamefont {J.}~\bibnamefont {Zeiher}}, \bibinfo {author}
  {\bibfnamefont {P.}~\bibnamefont {Schau{\ss}}}, \bibinfo {author}
  {\bibfnamefont {A.}~\bibnamefont {Rubio-Abadal}}, \bibinfo {author}
  {\bibfnamefont {T.}~\bibnamefont {Yefsah}}, \bibinfo {author} {\bibfnamefont
  {V.}~\bibnamefont {Khemani}}, \bibinfo {author} {\bibfnamefont {D.~A.}\
  \bibnamefont {Huse}}, \bibinfo {author} {\bibfnamefont {I.}~\bibnamefont
  {Bloch}}, \ and\ \bibinfo {author} {\bibfnamefont {C.}~\bibnamefont
  {Gross}},\ }\href {\doibase 10.1126/science.aaf8834} {\bibfield  {journal}
  {\bibinfo  {journal} {Science}\ }\textbf {\bibinfo {volume} {352}},\ \bibinfo
  {pages} {1547} (\bibinfo {year} {2016})}\BibitemShut {NoStop}%
\bibitem [{\citenamefont {Bordia}\ \emph {et~al.}(2017)\citenamefont {Bordia},
  \citenamefont {L\"uschen}, \citenamefont {Scherg}, \citenamefont
  {Gopalakrishnan}, \citenamefont {Knap}, \citenamefont {Schneider},\ and\
  \citenamefont {Bloch}}]{Bordia2017}%
  \BibitemOpen
  \bibfield  {author} {\bibinfo {author} {\bibfnamefont {P.}~\bibnamefont
  {Bordia}}, \bibinfo {author} {\bibfnamefont {H.}~\bibnamefont {L\"uschen}},
  \bibinfo {author} {\bibfnamefont {S.}~\bibnamefont {Scherg}}, \bibinfo
  {author} {\bibfnamefont {S.}~\bibnamefont {Gopalakrishnan}}, \bibinfo
  {author} {\bibfnamefont {M.}~\bibnamefont {Knap}}, \bibinfo {author}
  {\bibfnamefont {U.}~\bibnamefont {Schneider}}, \ and\ \bibinfo {author}
  {\bibfnamefont {I.}~\bibnamefont {Bloch}},\ }\href {\doibase
  10.1103/PhysRevX.7.041047} {\bibfield  {journal} {\bibinfo  {journal} {Phys.
  Rev. X}\ }\textbf {\bibinfo {volume} {7}},\ \bibinfo {pages} {041047}
  (\bibinfo {year} {2017})}\BibitemShut {NoStop}%
\bibitem [{\citenamefont {Hofstadter}(1976)}]{Hofstadter1976}%
  \BibitemOpen
  \bibfield  {author} {\bibinfo {author} {\bibfnamefont {D.~R.}\ \bibnamefont
  {Hofstadter}},\ }\href {\doibase 10.1103/PhysRevB.14.2239} {\bibfield
  {journal} {\bibinfo  {journal} {Phys. Rev. B}\ }\textbf {\bibinfo {volume}
  {14}},\ \bibinfo {pages} {2239} (\bibinfo {year} {1976})}\BibitemShut
  {NoStop}%
\bibitem [{\citenamefont {Thouless}\ \emph {et~al.}(1982)\citenamefont
  {Thouless}, \citenamefont {Kohmoto}, \citenamefont {Nightingale},\ and\
  \citenamefont {den Nijs}}]{Thouless1982}%
  \BibitemOpen
  \bibfield  {author} {\bibinfo {author} {\bibfnamefont {D.~J.}\ \bibnamefont
  {Thouless}}, \bibinfo {author} {\bibfnamefont {M.}~\bibnamefont {Kohmoto}},
  \bibinfo {author} {\bibfnamefont {M.~P.}\ \bibnamefont {Nightingale}}, \ and\
  \bibinfo {author} {\bibfnamefont {M.}~\bibnamefont {den Nijs}},\ }\href
  {\doibase 10.1103/PhysRevLett.49.405} {\bibfield  {journal} {\bibinfo
  {journal} {Phys. Rev. Lett.}\ }\textbf {\bibinfo {volume} {49}},\ \bibinfo
  {pages} {405} (\bibinfo {year} {1982})}\BibitemShut {NoStop}%
\bibitem [{\citenamefont {Ippoliti}\ \emph {et~al.}(2018)\citenamefont
  {Ippoliti}, \citenamefont {Geraedts},\ and\ \citenamefont
  {Bhatt}}]{Ippoliti2018}%
  \BibitemOpen
  \bibfield  {author} {\bibinfo {author} {\bibfnamefont {M.}~\bibnamefont
  {Ippoliti}}, \bibinfo {author} {\bibfnamefont {S.~D.}\ \bibnamefont
  {Geraedts}}, \ and\ \bibinfo {author} {\bibfnamefont {R.~N.}\ \bibnamefont
  {Bhatt}},\ }\href {\doibase 10.1103/PhysRevB.97.014205} {\bibfield  {journal}
  {\bibinfo  {journal} {Phys. Rev. B}\ }\textbf {\bibinfo {volume} {97}},\
  \bibinfo {pages} {014205} (\bibinfo {year} {2018})}\BibitemShut {NoStop}%
\bibitem [{\citenamefont {Krishna}\ \emph {et~al.}()\citenamefont {Krishna},
  \citenamefont {Ippoliti},\ and\ \citenamefont {Bhatt}}]{Krishna2018}%
  \BibitemOpen
  \bibfield  {author} {\bibinfo {author} {\bibfnamefont {A.}~\bibnamefont
  {Krishna}}, \bibinfo {author} {\bibfnamefont {M.}~\bibnamefont {Ippoliti}}, \
  and\ \bibinfo {author} {\bibfnamefont {R.~N.}\ \bibnamefont {Bhatt}},\
  }\href@noop {} {\bibinfo  {journal} {in preparation}\ }\BibitemShut {NoStop}%
\bibitem [{Note1()}]{Note1}%
  \BibitemOpen
\bibfield  {journal} {  }\bibinfo {note} {One recovers uncorrelated Gaussian
  white noise for $\sigma \to 0$. Setting $\sigma = 2 \ell _B$ improves the
  behavior of the single-particle localization length compared to $\sigma =0$
  (see \cite {Krishna2018} for details)}\BibitemShut {NoStop}%
\bibitem [{Note2()}]{Note2}%
  \BibitemOpen
  \bibinfo {note} {In hard disk terms, our ensemble corresponds to a density of
  0.4, much smaller than the close-packed density for hard disks in two
  dimensions (0.907).}\BibitemShut {Stop}%
\bibitem [{Note3()}]{Note3}%
  \BibitemOpen
  \bibinfo {note} {The degenerate subspace is the set of Slater determinants of
  orbitals in $\protect \ensuremath {\protect \mathcal {P}_{\protect \text
  {sb}}}$.}\BibitemShut {Stop}%
\bibitem [{\citenamefont {Atas}\ \emph {et~al.}(2013)\citenamefont {Atas},
  \citenamefont {Bogomolny}, \citenamefont {Giraud},\ and\ \citenamefont
  {Roux}}]{Atas2013}%
  \BibitemOpen
  \bibfield  {author} {\bibinfo {author} {\bibfnamefont {Y.~Y.}\ \bibnamefont
  {Atas}}, \bibinfo {author} {\bibfnamefont {E.}~\bibnamefont {Bogomolny}},
  \bibinfo {author} {\bibfnamefont {O.}~\bibnamefont {Giraud}}, \ and\ \bibinfo
  {author} {\bibfnamefont {G.}~\bibnamefont {Roux}},\ }\href {\doibase
  10.1103/PhysRevLett.110.084101} {\bibfield  {journal} {\bibinfo  {journal}
  {Phys. Rev. Lett.}\ }\textbf {\bibinfo {volume} {110}},\ \bibinfo {pages}
  {084101} (\bibinfo {year} {2013})}\BibitemShut {NoStop}%
\bibitem [{\citenamefont {Chayes}\ \emph {et~al.}(1986)\citenamefont {Chayes},
  \citenamefont {Chayes}, \citenamefont {Fisher},\ and\ \citenamefont
  {Spencer}}]{Chayes1986}%
  \BibitemOpen
  \bibfield  {author} {\bibinfo {author} {\bibfnamefont {J.~T.}\ \bibnamefont
  {Chayes}}, \bibinfo {author} {\bibfnamefont {L.}~\bibnamefont {Chayes}},
  \bibinfo {author} {\bibfnamefont {D.~S.}\ \bibnamefont {Fisher}}, \ and\
  \bibinfo {author} {\bibfnamefont {T.}~\bibnamefont {Spencer}},\ }\href
  {\doibase 10.1103/PhysRevLett.57.2999} {\bibfield  {journal} {\bibinfo
  {journal} {Phys. Rev. Lett.}\ }\textbf {\bibinfo {volume} {57}},\ \bibinfo
  {pages} {2999} (\bibinfo {year} {1986})}\BibitemShut {NoStop}%
\bibitem [{\citenamefont {{Chandran}}\ \emph {et~al.}(2015)\citenamefont
  {{Chandran}}, \citenamefont {{Laumann}},\ and\ \citenamefont
  {{Oganesyan}}}]{Chandran2015}%
  \BibitemOpen
  \bibfield  {author} {\bibinfo {author} {\bibfnamefont {A.}~\bibnamefont
  {{Chandran}}}, \bibinfo {author} {\bibfnamefont {C.~R.}\ \bibnamefont
  {{Laumann}}}, \ and\ \bibinfo {author} {\bibfnamefont {V.}~\bibnamefont
  {{Oganesyan}}},\ }\href@noop {} {\bibfield  {journal} {\bibinfo  {journal}
  {ArXiv e-prints}\ } (\bibinfo {year} {2015})},\ \Eprint
  {http://arxiv.org/abs/1509.04285} {arXiv:1509.04285 [cond-mat.dis-nn]}
  \BibitemShut {NoStop}%
\bibitem [{\citenamefont {Kj\"all}\ \emph {et~al.}(2014)\citenamefont
  {Kj\"all}, \citenamefont {Bardarson},\ and\ \citenamefont
  {Pollmann}}]{Kjall2014}%
  \BibitemOpen
  \bibfield  {author} {\bibinfo {author} {\bibfnamefont {J.~A.}\ \bibnamefont
  {Kj\"all}}, \bibinfo {author} {\bibfnamefont {J.~H.}\ \bibnamefont
  {Bardarson}}, \ and\ \bibinfo {author} {\bibfnamefont {F.}~\bibnamefont
  {Pollmann}},\ }\href {\doibase 10.1103/PhysRevLett.113.107204} {\bibfield
  {journal} {\bibinfo  {journal} {Phys. Rev. Lett.}\ }\textbf {\bibinfo
  {volume} {113}},\ \bibinfo {pages} {107204} (\bibinfo {year}
  {2014})}\BibitemShut {NoStop}%
\bibitem [{\citenamefont {Luitz}\ \emph {et~al.}(2015)\citenamefont {Luitz},
  \citenamefont {Laflorencie},\ and\ \citenamefont {Alet}}]{Luitz2015}%
  \BibitemOpen
  \bibfield  {author} {\bibinfo {author} {\bibfnamefont {D.~J.}\ \bibnamefont
  {Luitz}}, \bibinfo {author} {\bibfnamefont {N.}~\bibnamefont {Laflorencie}},
  \ and\ \bibinfo {author} {\bibfnamefont {F.}~\bibnamefont {Alet}},\ }\href
  {\doibase 10.1103/PhysRevB.91.081103} {\bibfield  {journal} {\bibinfo
  {journal} {Phys. Rev. B}\ }\textbf {\bibinfo {volume} {91}},\ \bibinfo
  {pages} {081103} (\bibinfo {year} {2015})}\BibitemShut {NoStop}%
\bibitem [{\citenamefont {Vosk}\ \emph {et~al.}(2015)\citenamefont {Vosk},
  \citenamefont {Huse},\ and\ \citenamefont {Altman}}]{Vosk2015}%
  \BibitemOpen
  \bibfield  {author} {\bibinfo {author} {\bibfnamefont {R.}~\bibnamefont
  {Vosk}}, \bibinfo {author} {\bibfnamefont {D.~A.}\ \bibnamefont {Huse}}, \
  and\ \bibinfo {author} {\bibfnamefont {E.}~\bibnamefont {Altman}},\ }\href
  {\doibase 10.1103/PhysRevX.5.031032} {\bibfield  {journal} {\bibinfo
  {journal} {Phys. Rev. X}\ }\textbf {\bibinfo {volume} {5}},\ \bibinfo {pages}
  {031032} (\bibinfo {year} {2015})}\BibitemShut {NoStop}%
\bibitem [{\citenamefont {Potter}\ \emph {et~al.}(2015)\citenamefont {Potter},
  \citenamefont {Vasseur},\ and\ \citenamefont {Parameswaran}}]{Potter2015}%
  \BibitemOpen
  \bibfield  {author} {\bibinfo {author} {\bibfnamefont {A.~C.}\ \bibnamefont
  {Potter}}, \bibinfo {author} {\bibfnamefont {R.}~\bibnamefont {Vasseur}}, \
  and\ \bibinfo {author} {\bibfnamefont {S.~A.}\ \bibnamefont {Parameswaran}},\
  }\href {\doibase 10.1103/PhysRevX.5.031033} {\bibfield  {journal} {\bibinfo
  {journal} {Phys. Rev. X}\ }\textbf {\bibinfo {volume} {5}},\ \bibinfo {pages}
  {031033} (\bibinfo {year} {2015})}\BibitemShut {NoStop}%
\bibitem [{\citenamefont {Dumitrescu}\ \emph {et~al.}(2017)\citenamefont
  {Dumitrescu}, \citenamefont {Vasseur},\ and\ \citenamefont
  {Potter}}]{Dumitrescu2017}%
  \BibitemOpen
  \bibfield  {author} {\bibinfo {author} {\bibfnamefont {P.~T.}\ \bibnamefont
  {Dumitrescu}}, \bibinfo {author} {\bibfnamefont {R.}~\bibnamefont {Vasseur}},
  \ and\ \bibinfo {author} {\bibfnamefont {A.~C.}\ \bibnamefont {Potter}},\
  }\href {\doibase 10.1103/PhysRevLett.119.110604} {\bibfield  {journal}
  {\bibinfo  {journal} {Phys. Rev. Lett.}\ }\textbf {\bibinfo {volume} {119}},\
  \bibinfo {pages} {110604} (\bibinfo {year} {2017})}\BibitemShut {NoStop}%
\bibitem [{\citenamefont {Khemani}\ \emph {et~al.}(2017)\citenamefont
  {Khemani}, \citenamefont {Lim}, \citenamefont {Sheng},\ and\ \citenamefont
  {Huse}}]{Khemani2017}%
  \BibitemOpen
  \bibfield  {author} {\bibinfo {author} {\bibfnamefont {V.}~\bibnamefont
  {Khemani}}, \bibinfo {author} {\bibfnamefont {S.~P.}\ \bibnamefont {Lim}},
  \bibinfo {author} {\bibfnamefont {D.~N.}\ \bibnamefont {Sheng}}, \ and\
  \bibinfo {author} {\bibfnamefont {D.~A.}\ \bibnamefont {Huse}},\ }\href
  {\doibase 10.1103/PhysRevX.7.021013} {\bibfield  {journal} {\bibinfo
  {journal} {Phys. Rev. X}\ }\textbf {\bibinfo {volume} {7}},\ \bibinfo {pages}
  {021013} (\bibinfo {year} {2017})}\BibitemShut {NoStop}%
\end{thebibliography}%

\end{document}